\newcommand{\diracslash}[1]{#1\llap{/\kern2pt}}

\newcommand{\be}{\begin{equation}}
\newcommand{\ee}{\end{equation}}
\newcommand{\bea}{\begin{eqnarray}}
\newcommand{\eea}{\end{eqnarray}}
\newcommand{\ba}[1]{\begin{array}{#1}}
\newcommand{\ea}{\end{array}}

\documentclass[preprint,prd,aps,floats,nofootinbib,showpacs,floatfix]{revtex4}
\usepackage{graphicx}
\begin{document}

\title{Kaon properties in (proto-)neutron star matter}

\author{Amruta Mishra}
\email{amruta@physics.iitd.ac.in}
\affiliation{Department of Physics,Indian Institute of Technology,Delhi,
New Delhi - 110 016, India}

\author{Arvind Kumar}
\email{iitd.arvind@gmail.com}
\affiliation{Department of Physics,Indian Institute of Technology,Delhi,
New Delhi - 110 016, India}

\author{Sambuddha Sanyal}
\email{sambuddhasanyal@gmail.com}
\affiliation{Department of Physics,Indian Institute of Technology,Delhi,
New Delhi - 110 016, India}

\author{V. Dexheimer}
\email{dexheimer@th.physik.uni-frankfurt.de}
\affiliation{
Frankfurt Institute for Advanced Studies,
J.W. Goethe Universit\"at,
Ruth-Moufang-Str. 1, D-60438 Frankfurt am Main, Germany}

\author{Stefan Schramm}
\email{schramm@th.physik.uni-frankfurt.de}
\affiliation{
Frankfurt Institute for Advanced Studies,
J.W. Goethe Universit\"at,
Ruth-Moufang-Str. 1, D-60438 Frankfurt am Main, Germany}

\begin{abstract}
The modification on kaon and antikaon properties of in the interior of 
(proto-)neutron stars is investigated using a chiral SU(3) model. The 
parameters of the model are fitted to nuclear matter saturation properties, 
baryon octet vacuum masses, hyperon optical potentials and low energy a
kaon-nucleon scattering lengths. We study the kaon/antikaon medium 
modification and explore the possibility of antikaon condensation in 
(proto-)neutron star matter at zero as well as finite temperature/entropy 
and neutrino content. The effect of hyperons on kaon and antikaon optical
potentials is also investigated at different stages of the neutron star 
evolution.

\end{abstract}


\maketitle

\section{Introduction}

The study of kaon and antikaon properties is relevant for neutron star 
phenomenology as well as relativistic heavy-ion collision experiments. 
It was first suggested by Kaplan and Nelson \cite{kaplan} that the drop 
in the mass of antikaons in nuclear medium, arising from an attractive 
antikaon-nucleon interaction, might lead to Bose-Einstein condensation 
in the interior of neutron stars. Since then, a lot of work has been done 
in the topic of antikaon condensation in compact stars \cite{kaoncond}. 
The s-wave $K^-$ condensation sets in when the in-medium energy of $K^-$
at zero momentum equals the chemical potential of $K^-$. The effect of 
the $K^-$ condensate is to replace the electrons in maintaining 
the charge neutrality condition. The formation of $\bar {K^0}$ condensate 
in neutron stars in a mean field approach has also been investigated 
\cite{deba}. The condition for the onset of  neutral antikaon $\bar {K^0}$ 
condensation is $\omega_{\bar {K^0}}(k=0)=0$.

The gross properties of (proto-)neutron stars depend sensitively 
on the equation of state of dense, electric charge neutral matter. 
On the other hand, recent neutron star observations impose constraints 
on possible nuclear matter equations of state (EOS's). As a result, 
nuclear matter EOS's obtained using effective models should be consistent 
with astrophysical bounds in order to be used for neutron star matter 
calculations \cite{blaschkeozel}. Recently, nuclear matter EOS's have 
also been investigated consistently with collective flow data from heavy 
ion collision experiments \cite{klahn}. The in-medium modification of 
kaon/antikaon properties have been studied using effective hadronic models 
based on meson exchange \cite{kaoneff}, chiral perturbation theory 
(using data from kaonic atom \cite{kaonicatom}) as well as using coupled 
channel approach \cite{cplch}. These in-medium properties can be observed 
experimentally in relativistic heavy ion collisions \cite{kaonhic} 
as well as in neutron star phenomenology 
\cite{kaplan,kaoncond,deba}. Due to charge neutrality, the bulk matter 
in proto-neutron stars is isospin asymmetric. The isospin effects in 
hot and dense hadronic matter (already investigated in \cite{asym}) 
are thus important in the context of isospin asymmetric heavy-ion 
collision experiments \cite{asymexpt} as well as for matter in the 
interior of neutron stars.

In the present investigation we use a chiral SU(3) model \cite{paper3,kristof1}
for the description of matter inside (proto-)neutron stars. The kaon (antikaon)
energies, modified in the medium due to their interaction with nucleons, 
were also studied within this framework \cite{kmeson1,isoamss} consistently
with the low energy KN scattering data \cite{juergen,cohen}. In the present 
work, we investigate the kaon and antikaon optical potentials in the asymmetric
nucleonic/hyperonic matter. We also study in detail the possibility of kaon 
condensation on different stages of the neutron star cooling. These stages 
are defined by finite temperature/entropy and neutrino content. 

The outline of the paper is as follows: In section II we briefly review 
the SU(3) model used in the present investigation. Section III describes 
the medium modification of the K($\bar K$) mesons in this effective model. 
In section IV, we discuss the results obtained for the optical potentials 
of kaons and antikaons in (proto-)neutron stars. Section V summarizes our 
results and presents the conclusions.

\section{The hadronic chiral $SU(3) \times SU(3)$ model}
In this section the various terms of the effective hadronic Lagrangian density used
\be
{\cal L} = {\cal L}_{kin} + {\cal L}_{int}+ {\cal L}_{vec} + {\cal L}_{scal} + {\cal L}_{SB},
\label{genlag} \ee
are discussed. This expression derives from a relativistic quantum field 
theoretical model of baryons and mesons constructed adopting a nonlinear 
realization of chiral symmetry \cite{weinberg,coleman,bardeen} and broken 
scale invariance (for details see \cite{paper3,kristof1,hartree}) to describe 
strongly interacting nuclear matter. The chiral $SU(3) \times SU(3)$ model 
was already successfully used to describe nuclear matter, finite nuclei, 
hypernuclei and neutron stars. The Lagrangian density contains the baryon 
octet, the spin-0 and spin-1 meson multiplets as elementary degrees of 
freedom. In Eq. \ref{genlag}, $ {\cal L}_{kin} $ is the kinetic energy 
term and $  {\cal L}_{int}  $ contains the baryon-meson interactions in
 which the baryon-spin-0 mesons generate the baryon masses. The term 
$ {\cal L}_{vec} $ describes the dynamical mass generation of the vector 
mesons and contains additionally quartic self-interactions of the vector 
fields. The term  ${\cal L}_{scal} $ contains the scalar meson self-interaction
 terms, that induce spontaneous breaking of chiral symmetry, as well as a scale
 invariance breaking logarithmic potential. Finally, the term $ {\cal L}_{SB} $
 describes the explicit chiral symmetry breaking.

The baryon-scalar meson interactions generate the baryon masses through 
the coupling of the baryons to the non-strange $ \sigma (\sim
\langle\bar{u}u + \bar{d}d\rangle) $ and the strange 
$ \zeta(\sim\langle\bar{s}s\rangle) $ scalar quark condensates. 
The parameters corresponding to these interactions are adjusted to fix 
the baryon masses to their measured vacuum values. It should be emphasized 
that the nucleon mass also depends on the {\em strange condensate} $ \zeta $.
 In analogy to the baryon-scalar meson coupling there exist two independent 
baryon-vector meson interaction terms corresponding to the antisymmetric 
and symmetric couplings. Here we use the antisymmetric coupling 
\cite{paper3,isoamss} following the universality principle  \cite{saku69} 
and the vector meson dominance model which shows that the symmetric coupling 
should be small. Additionally we choose parameters so as to decouple the 
strange vector field $ \phi_\mu\sim\bar{s} \gamma_\mu s $ from the nucleons 
\cite{paper3,isoamss}, that corresponds to an ideal mixing between $\omega$ 
and $\phi$. A small deviation of the mixing angle from the ideal mixing 
(used in \cite {dumbrajs,rijken,hohler1}) has not been taken into account 
in the present investigation.

The Lagrangian density terms corresponding to the self-interaction for 
the vector mesons ${\cal L}_{vec}$, scalar mesons ${\cal L}_{scal}$ and 
the one corresponding to the explicit chiral symmetry breaking ${\cal L}_{SB}$ 
have been described in detail in references \cite{paper3,isoamss}. For the 
non linear interaction of vector mesons, we use the invariant which does not
 generate the $\rho$-$\omega$ coupling. That is in general agreement with the
 observed small mixing between the two mesons \cite{dexschr8} and allows for 
more massive neutron stars.

To investigate the hadronic properties in the medium, we write the Lagrangian 
density within the chiral SU(3) model in the mean field approximation 
\cite{walecka}. With this we can determine the expectation values of the meson fields by minimizing the thermodynamical potential \cite{kristof1,hartree}.

\section{Kaon (antikaon) interactions in the chiral SU(3) model}

In this section, we derive the dispersion relations for the $K (\bar K)$ 
\cite{kmeson} and calculate their optical potentials in asymmetric hadronic 
matter \cite{isoamss}. In the present model, the interactions of kaons and 
antikaons with scalar fields $\sigma$ (non-strange) and $\zeta$ (strange) 
and the scalar-isovector field $\delta$ as well as a vector interaction 
with the nucleons (the so--called Weinberg-Tomozawa interaction)
modify the energy of ${\rm K}(\bar {\rm  K})$ mesons in the medium. 
It might be noted here that the interaction of pseudoscalar mesons 
with the vector mesons, in addition to the pseudoscalar meson--nucleon 
vector interaction, leads to double counting in the linear realization 
of the chiral effective theory \cite{borasoy}. Within the nonlinear 
realization of chiral effective theories, such an interaction does not 
arise in the leading or sub-leading order, but only as a higher order 
contribution \cite{borasoy}. Hence, the vector meson-pseudoscalar meson 
interaction is not considered within the present investigation.

In the following, we derive the dispersion relations for kaons and antikaons
and study the dependence of kaon and antikaon optical potentials on isospin 
asymmetry. In order to do this, we include effects of isospin asymmetry 
originating from the scalar-isovector $\delta$ field as well as a vector 
interaction with nucleons \cite{isoamss}. In addition, in the present 
investigation we consider an isospin dependent range term arising from 
the interaction with nucleons, which was not taken into account in 
Ref. \cite{isoamss}.

The term in the Lagrangian density that represents the interaction between 
baryons and kaons modifies the energies of the $K(\bar K)$-mesons,
and is given by
\begin{eqnarray}
\cal L _{KB} & = & -\frac {i}{4 f_K^2} \Big [\Big ( 2 \bar p \gamma^\mu p
+\bar n \gamma ^\mu n -\bar {\Sigma^-}\gamma ^\mu \Sigma ^-
+\bar {\Sigma^+}\gamma ^\mu \Sigma ^+
- 2\bar {\Xi^-}\gamma ^\mu \Xi ^-
- \bar {\Xi^0}\gamma ^\mu \Xi^0 \Big)
\nonumber \\
& \times &
\Big(K^- (\partial_\mu K^+) - (\partial_\mu {K^-})  K^+ \Big )
\nonumber \\
& + &
\Big ( \bar p \gamma^\mu p
+ 2\bar n \gamma ^\mu n +\bar {\Sigma^-}\gamma ^\mu \Sigma ^-
-\bar {\Sigma^+}\gamma ^\mu \Sigma ^+
- \bar {\Xi^-}\gamma ^\mu \Xi ^-
- 2 \bar {\Xi^0}\gamma ^\mu \Xi^0 \Big)
\nonumber \\
& \times &
\Big(\bar {K^0} (\partial_\mu K^0) - (\partial_\mu {\bar {K^0}})  K^0 \Big )
\Big ]
\nonumber \\
 &+ & \frac{m_K^2}{2f_K} \Big [ (\sigma +\sqrt 2 \zeta+\delta)(K^+ K^-)
 + (\sigma +\sqrt 2 \zeta-\delta)(K^0 \bar { K^0})
\Big ] \nonumber \\
& - & \frac {1}{f_K}\Big [ (\sigma +\sqrt 2 \zeta +\delta)
(\partial _\mu {K^+})(\partial ^\mu {K^-})
+(\sigma +\sqrt 2 \zeta -\delta)
(\partial _\mu {K^0})(\partial ^\mu \bar {K^0})
\Big ]
\nonumber \\
&+ & \frac {d_1}{2 f_K^2}(\bar p p +\bar n n +\bar {\Lambda^0}{\Lambda^0}
+\bar {\Sigma ^+}{\Sigma ^+}
+\bar {\Sigma ^0}{\Sigma ^0}
+\bar {\Sigma ^-}{\Sigma ^-}
+\bar {\Xi ^-}{\Xi ^-}
+\bar {\Xi ^0}{\Xi ^0}
 )
(\partial _\mu {\bar K})(\partial ^\mu K)\nonumber \\
&+& \frac {d_2}{2 f_K^2} \Big [
(\bar p p+\frac {5}{6} \bar {\Lambda^0}{\Lambda^0}
+\frac {1}{2} \bar {\Sigma^0}{\Sigma^0}
+\bar {\Sigma^+}{\Sigma^+}
+\bar {\Xi^-}{\Xi^-}
+\bar {\Xi^0}{\Xi^0}
) (\partial_\mu K^+)(\partial^\mu K^-) 
\nonumber \\
 &+ &(\bar n n
+\frac {5}{6} \bar {\Lambda^0}{\Lambda^0}
+\frac {1}{2} \bar {\Sigma^0}{\Sigma^0}
+\bar {\Sigma^-}{\Sigma^-}
+\bar {\Xi^-}{\Xi^-}
+\bar {\Xi^0}{\Xi^0}
) (\partial_\mu K^0)(\partial^\mu {\bar {K^0}})
\Big ].
\label{lagd}
\end{eqnarray}
In Eq. \ref{lagd} the first line stands for the vector interaction term 
(Weinberg-Tomozawa term) obtained from the kinetic term of the Lagrangian 
density \cite{isoamss}. The second term, which gives an attractive 
interaction for the $K$-mesons, is obtained from the explicit symmetry 
breaking term \cite{kmeson1,isoamss}. The third term arises within 
the present chiral model from the kinetic term of the pseudoscalar 
mesons \cite{isoamss}. The fourth and fifth terms in Eq. \ref{lagd} 
arise from
\begin{equation}
{\cal L }_{(d_1)}^{BM} =\frac {d_1}{2} Tr (u_\mu u ^\mu)Tr( \bar B B),
\label{d1ld}
\end{equation}
and
\begin{equation}
{\cal L }_{(d_2)}^{BM} =d_2 Tr (\bar B u_\mu u ^\mu B),
\label{d2ld}
\end{equation}
according to Ref. \cite{kmeson1,isoamss}. The last three terms in 
Eq. \ref{lagd} represent the range term with the last part being 
the isospin asymmetric interaction. The Fourier transformation of 
the equation-of-motion for kaons (antikaons) leads to the dispersion 
relation 
\begin{equation}
-\omega^2+ {\vec k}^2 + m_K^2 -\Pi(\omega, |\vec k|,\rho)=0,
\label{dispk}
\end{equation}
where $\Pi$ denotes the kaon (antikaon) self energy in the medium.

Explicitly, the self energy $\Pi (\omega,|\vec k|)$ for the kaon doublet ($K^+$,$K^0$) arising from the interaction (Eq. \ref{lagd}) is given by
\begin{eqnarray}
\Pi (\omega, |\vec k|) &= & -\frac {1}{4 f_K^2}\Big [3 (\rho_p +\rho_n)
\pm (\rho_p -\rho_n) \pm 2 (\rho_{\Sigma^+}-\rho_{\Sigma^-})
-\big ( 3 (\rho_{\Xi^-} +\rho_{\Xi^0}) \pm (\rho_{\Xi^-} -\rho_{\Xi^0})
\big)
\Big ] \omega\nonumber \\
&+&\frac {m_K^2}{2 f_K} (\sigma ' +\sqrt 2 \zeta ' \pm \delta ')
\nonumber \\ & +& \Big [- \frac {1}{f_K}
(\sigma ' +\sqrt 2 \zeta ' \pm \delta ')
+\frac {d_1}{2 f_K ^2} (\rho_s ^p +\rho_s ^n
+{\rho^s} _{\Lambda^0}+{\rho^s} _{\Sigma^+}+{\rho^s} _{\Sigma^0}
+{\rho^s} _{\Sigma^-} +{\rho^s} _{\Xi^-} +{\rho^s} _{\Xi^0}
)\nonumber \\
&+&\frac {d_2}{4 f_K ^2} \Big (({\rho^s} _p +{\rho^s} _n)
\pm   ({\rho^s} _p -{\rho^s} _n)
+{\rho^s} _{\Sigma ^0}+\frac {5}{3} {\rho^s} _{\Lambda^0}
+ ({\rho^s} _{\Sigma ^+}+{\rho^s} _{\Sigma ^-})
\pm ({\rho^s} _{\Sigma ^+}-{\rho^s} _{\Sigma ^-})\nonumber \\
 &+ & 2 {\rho^s} _ {\Xi^-}+
2 {\rho^s} _ {\Xi^0}
\Big )
\Big ]
(\omega ^2 - {\vec k}^2),
\label{selfk}
\end{eqnarray}
where the $\pm$ signs refer to $K^+$ and $K^0$, respectively. In the equation above $\sigma'(=\sigma-\sigma _0)$, $\zeta'(=\zeta-\zeta_0)$ and  $\delta'(=\delta-\delta_0)$ are the fluctuations of the scalar-isoscalar fields $\sigma$ and $\zeta$, and the third component of the scalar-isovector field $\delta$ from their vacuum expectation values. The vacuum expectation value of $\delta$ is zero ($\delta_0$=0) because otherwise the vacuum isospin symmetry would be broken. It is important to note that in the present work the small isospin breaking effect coming from the mass and charge difference of the up and down quarks has been neglected. The variables $\rho_i$ and ${\rho^s}_{i}$ stand for number density and scalar density of baryon type $i$, with $i=p,n, \Lambda, \Sigma ^\pm, \Sigma ^0, \Xi ^-, \Xi ^0$. Similarly, the self-energy for the antikaon doublet ($K^-$,$\bar {K^0}$) is calculated as
\begin{eqnarray}
\Pi (\omega, |\vec k|) &= & \frac {1}{4 f_K^2}\Big [3 (\rho_p +\rho_n)
\pm (\rho_p -\rho_n) \pm 2 (\rho_{\Sigma^+}-\rho_{\Sigma^-})
- \big ( 3 (\rho_{\Xi^-} +\rho_{\Xi^0}) \pm (\rho_{\Xi^-} -\rho_{\Xi^0})
\big)
\Big ] \omega\nonumber \\
&+&\frac {m_K^2}{2 f_K} (\sigma ' +\sqrt 2 \zeta ' \pm \delta ')
\nonumber \\ & +& \Big [- \frac {1}{f_K}
(\sigma ' +\sqrt 2 \zeta ' \pm \delta ')
+\frac {d_1}{2 f_K ^2} (\rho_s ^p +\rho_s ^n
+{\rho^s} _{\Lambda^0}+{\rho^s} _{\Sigma^+}+{\rho^s} _{\Sigma^0}
+{\rho^s} _{\Sigma^-} +{\rho^s} _{\Xi^-} +{\rho^s} _{\Xi^0}
)\nonumber \\
&+&\frac {d_2}{4 f_K ^2} \Big (({\rho^s} _p +{\rho^s} _n)
\pm   ({\rho^s} _p -{\rho^s} _n)
+{\rho^s} _{\Sigma ^0}+\frac {5}{3} {\rho^s} _{\Lambda^0}
+ ({\rho^s} _{\Sigma ^+}+{\rho^s} _{\Sigma ^-})
\pm ({\rho^s} _{\Sigma ^+}-{\rho^s} _{\Sigma ^-})\nonumber \\
 &+ & 2 {\rho^s} _ {\Xi^-}+ 2 {\rho^s} _ {\Xi^0}
\Big )
\Big ]
(\omega ^2 - {\vec k}^2),
\label{selfkb}
\end{eqnarray}
where the $\pm$ signs refer to $K^-$ and $\bar {K^0}$, respectively.

The optical potentials are calculated from kaon and antikaon energies using
\be
U(\omega, k) = \omega (k) -\sqrt {k^2 + m_K ^2},
\ee
where $m_K$ is the kaon (antikaon) vacuum mass. The parameters $d_1$ and $d_2$ are calculated from empirical values of the KN scattering lengths
for I=0 and I=1 channels. These are taken to be (as in \cite{thorsson,juergen,barnes})
\be
a _{KN} (I=0) \approx -0.09 ~ {\rm {fm}},\;\;\;\;
a _{KN} (I=1) \approx -0.31 ~ {\rm {fm}},
\label{akn01emp}
\ee
leading to the isospin averaged KN scattering length
\be
\bar a _{KN}=\frac {1}{4} a_{KN}(I=0)+
\frac {3}{4} a_{KN}(I=1) \approx -0.255 ~ \rm {fm}.
\label{aknemp}
\ee

\begin{figure}
\includegraphics[width=16cm,height=16cm]{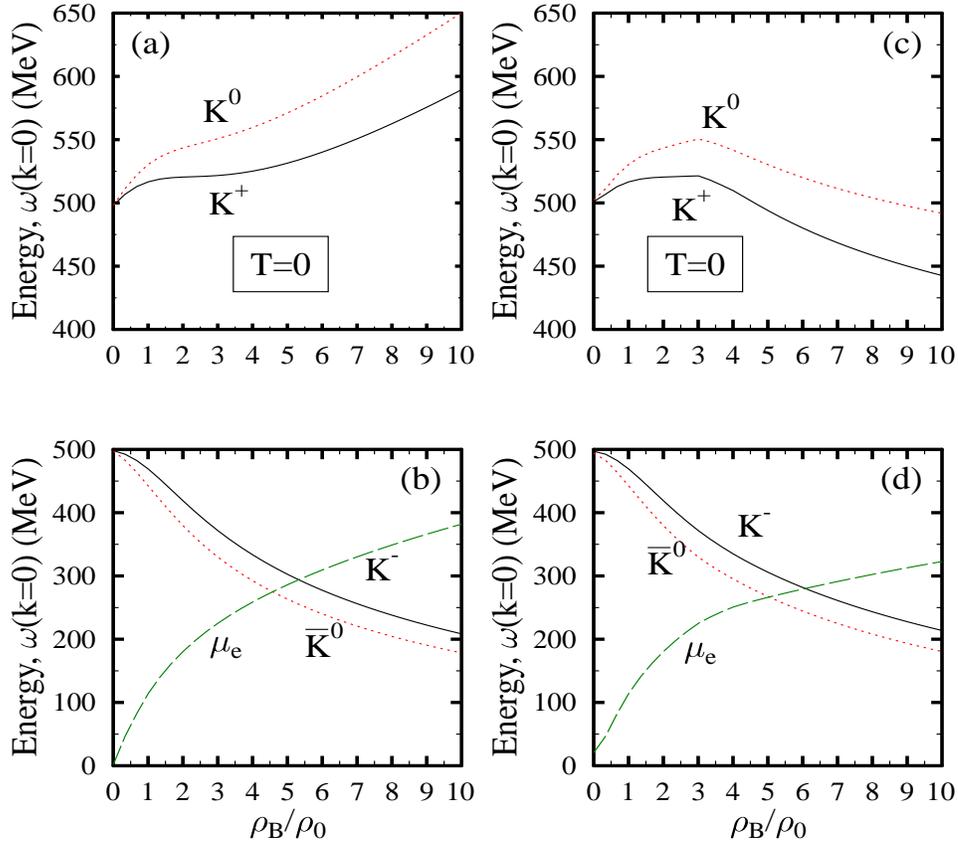}
\caption{
Kaon and antikaon energies (at zero momentum) as well as electron chemical 
potentials plotted for nucleonic matter in (a) and (b) and hyperonic matter
in (c) and (d) at T=0 and in beta equilibrium. 
}
\label{T0nm}
\end{figure}

\begin{figure}
\includegraphics[width=16cm,height=16cm]{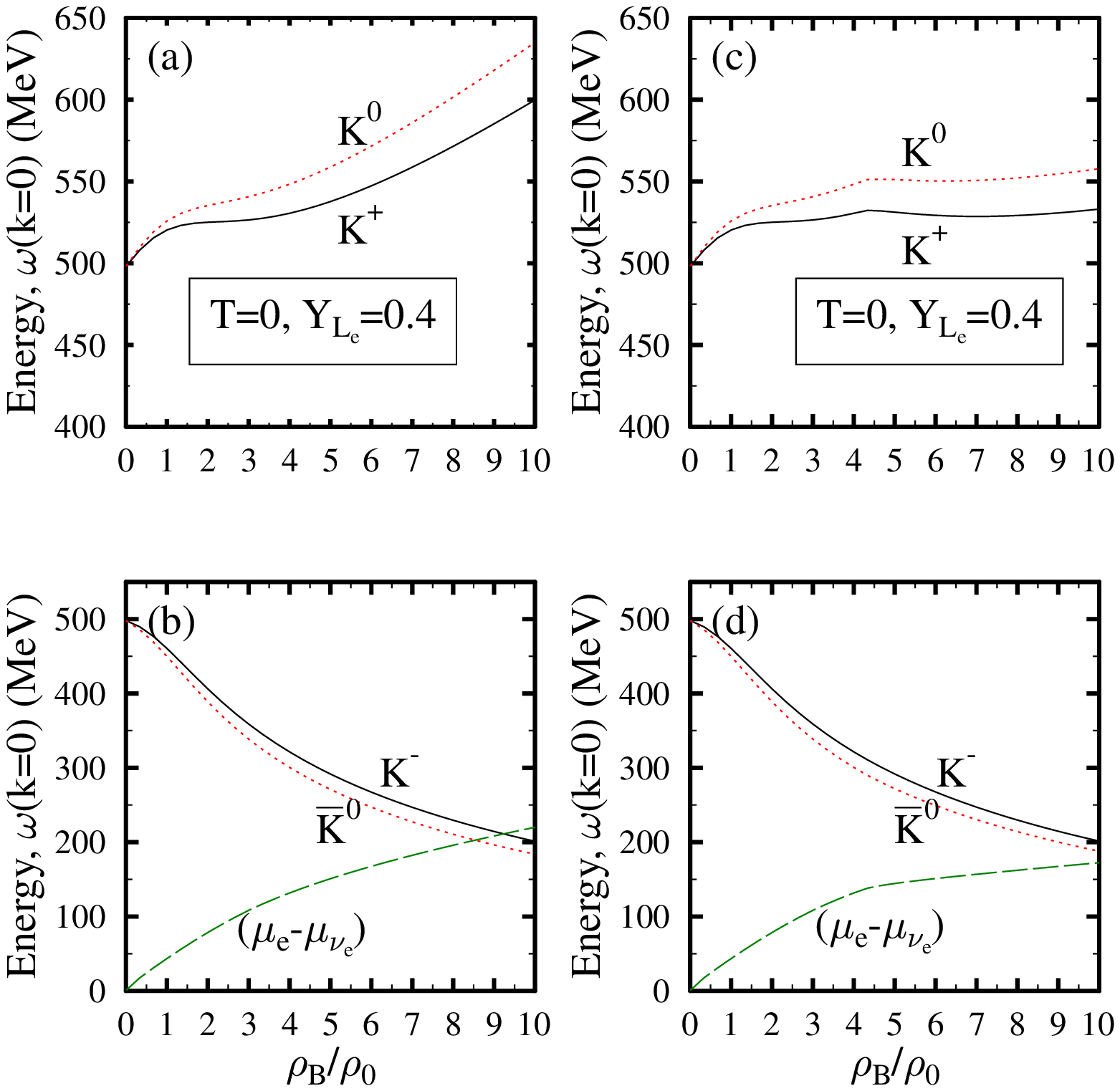}
\caption{
Kaon and antikaon energies (at zero momentum) as well as electron chemical 
potentials plotted for nucleonic matter in (a) and (b) and hyperonic matter 
in (c) and (d) at T=0 and with $Y_{L_e}$=0.4.
}
\end{figure}

\begin{figure}
\includegraphics[width=16cm,height=16cm]{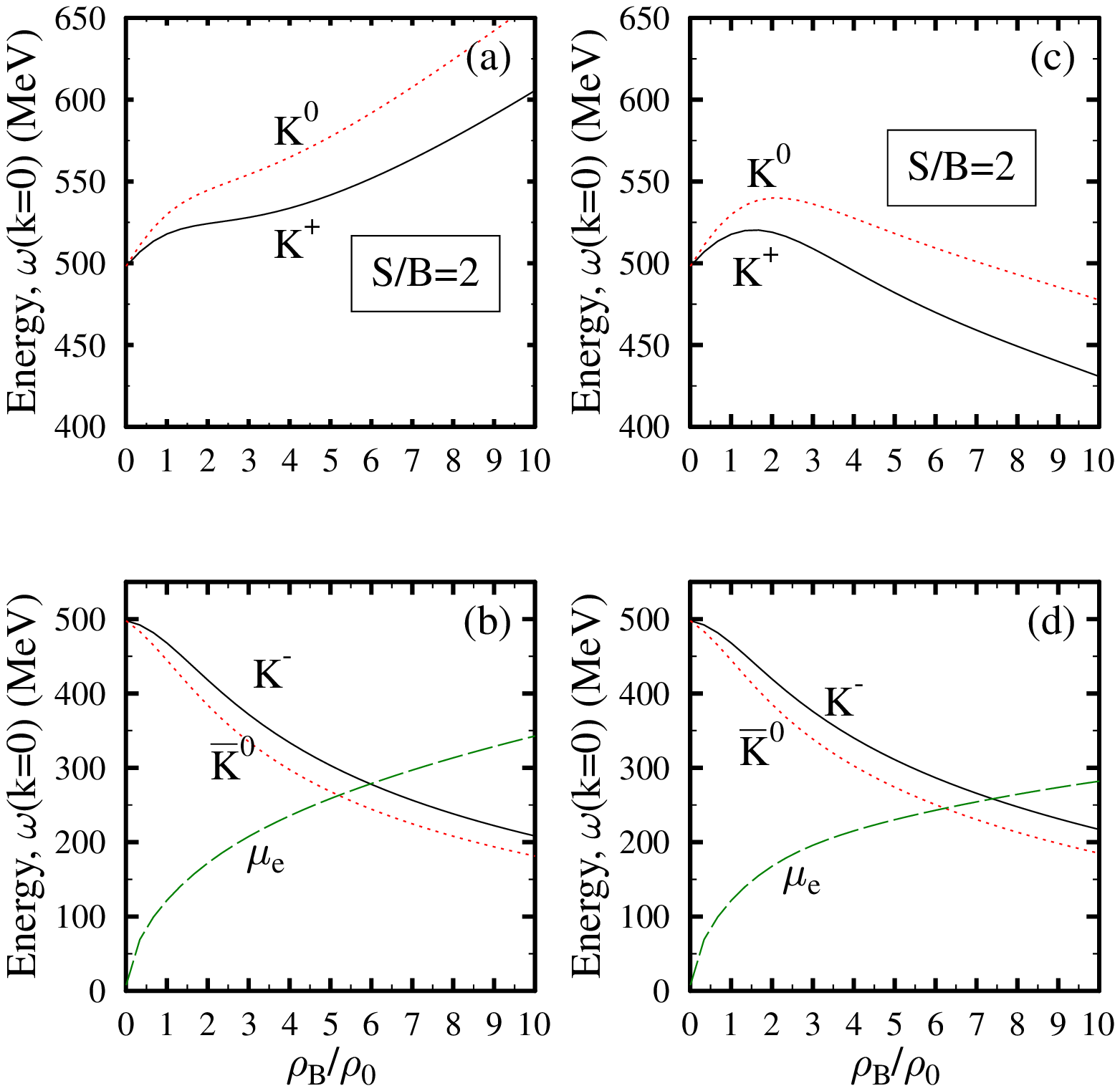}
\caption{
Kaon and antikaon energies (at zero momentum) as well as electron chemical
potentials plotted for nucleonic matter in (a) and (b) and hyperonic matter
 in (c) and (d) with $S/B=2$  and in beta equilibrium.
}
\end{figure}

\begin{figure}
\includegraphics[width=16cm,height=16cm]{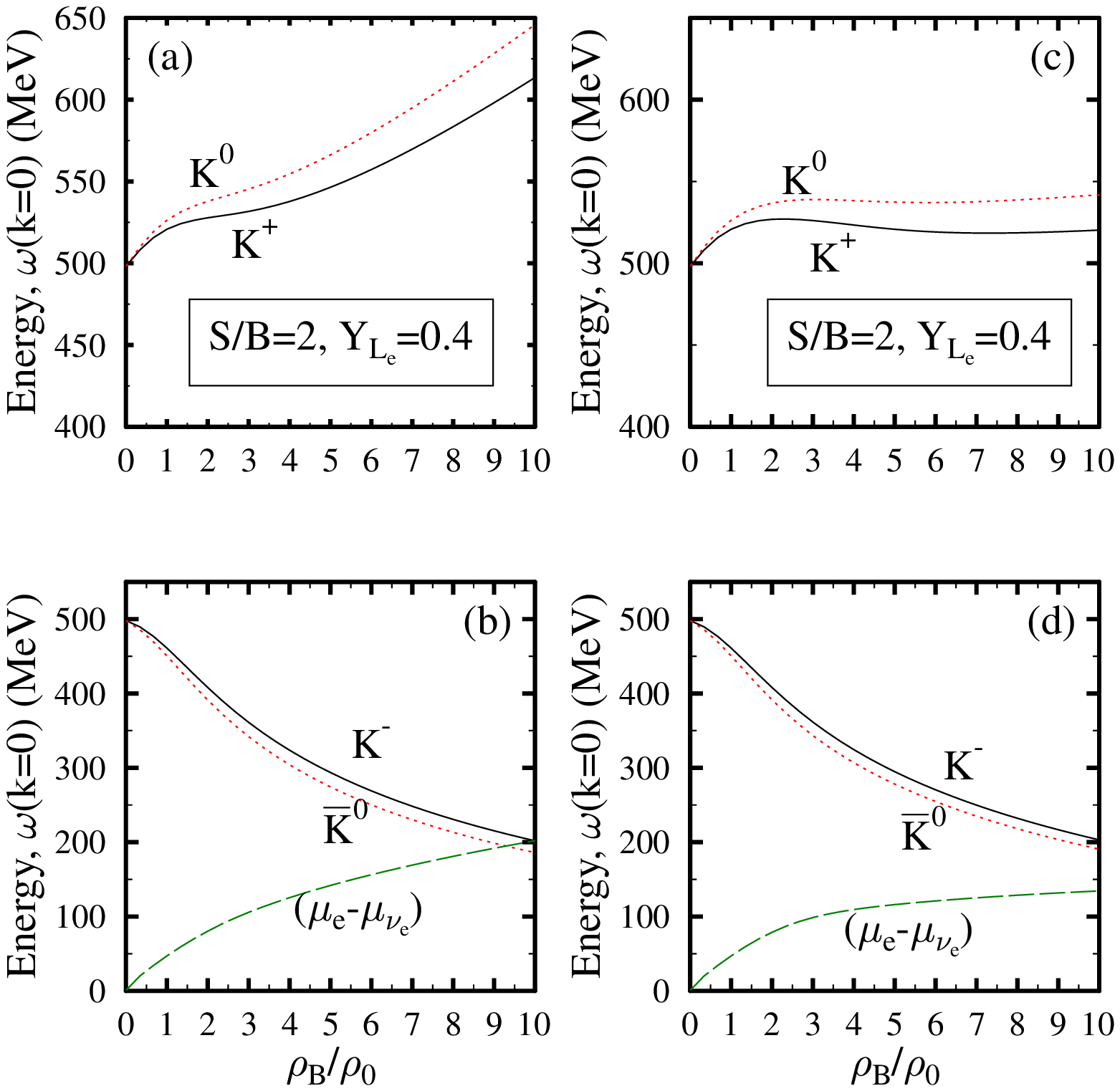}
\caption{
Kaon and antikaon energies (at zero momentum) as well as electron chemical potentials plotted for nucleonic matter in (a) and (b) and hyperonic matter in (c) and (d) with $S/B=2$  and $Y_{L_e}$=0.4.
}
\end{figure}

\section{Results and Discussions}
\label{kmass}

The results for the numerical analysis of the presence of kaons and antikaons 
in (proto-)neutron star matter are presented in this section. Using a chiral
 SU(3) model, the kaon and antikaon energies are investigated in the electric 
charge neutral, isospin asymmetric nucleonic matter as well as in hyperonic
 matter. The effects of finite temperature/entropy as well as neutrino trapping
 on the in-medium kaon (antikaon) masses and the possibility of antikaon 
condensation are studied under these circumstances. In the present calculation,
 the model parameters are determined by fitting the baryon vacuum  masses,
 nuclear matter saturation properties and hyperon optical potentials. 
The coefficients $d_1$ and $d_2$, calculated from the empirical values 
of KN scattering lengths for I=0 and I=1 channels (Eq. \ref{akn01emp}), 
are $2.5/{m_K}$ and $0.72/{m_K}$ respectively.

In the interior of (proto-)neutron stars, the weak interaction processes 
$n \rightarrow p +e^-+\bar {\nu_e}$ and $e^-\rightarrow K^-+\nu_e$ can 
occur. Assuming these processes to be in chemical equilibrium, we have 
\be
\mu_n-\mu_p=\mu_e-\mu_{\nu_e}=\mu_{K^-}.
\ee
For the onset of $K^-$ condensation $\mu _{K^-}=\omega_{K^-}$.

In Fig. 1, we plot the kaon and antikaon energies at zero 
momentum in (a) and (b) as functions of the baryon density for charge 
neutral and beta equilibrated nucleonic matter at zero temperature. 
The modifications to these energies are also shown when hyperons are 
included in the dense matter in (c) and (d) in the same figure. 
The electron chemical potential plotted as a function of density in 
(b) and (d) show whether there is a possibility of antikaon condensation 
in the nucleonic/hyperonic matter. For nucleonic matter, the onset 
for $K^-$ condensation takes place at a density of about 5.3 $\rho_0$. 
The net effect of the $K^-$ condensation is that the respective meson 
takes the role of electrons maintaining charge neutrality. The inclusion 
of hyperons shifts the condensation threshold density to 6.1 $\rho_0$, 
a higher value compared to the case that doesn't take hyperons into 
account (as in \cite{kaoncond,deba}). This is due to the fact that 
in the presence of hyperons, the negatively charged baryons take 
the role of electrons in maintaining charge neutrality. Consequently 
the value of the electron chemical potential is reduced, thus shifting 
the threshold density for onset of $K^-$ condensation to a higher value 
as compared to the threshold density for nucleonic matter. As a result, 
in the presence of hyperons the $K^-$ condensation takes place more 
towards the center of the neutron star.

Right after the supernova explosion, the proto-neutron star has 
a high amount of neutrinos. They are included in the calculation 
with a chemical potential $\mu_{\nu_e}$ by fixing the lepton number 
defined as $Y_{L_e}=(\rho_e+\rho_{\nu_e})/\rho_B$, where $\rho_B$ 
is the total baryonic number density. In this case, there is not 
only a large number of neutrinos in the star, but also an increased 
electron density. Therefore, demanding charge neutrality, the proton 
number density increases and with higher proton density the bulk matter 
in the star becomes more isospin symmetric, leading to a decrease in
 the neutron Fermi energy. Thus increasing lepton number softens the 
equation of state (EOS) and consequently leads to smaller star maximum 
masses. In Fig. 2 we plot $K$ and $\bar K$ energies at zero momentum 
in charge neutral matter with finite lepton number fraction $Y_{L_e}$=0.4 
and zero temperature. We see that the threshold density for the onset 
of $K^-$ condensation is shifted to higher values in the case with 
neutrino trapping compared to the neutrino free one (in agreement with 
Ref. \cite{kaoncond,deba}). For nucleonic matter, our model predicts 
$K^-$ condensation at a density of around 9.3 $\rho_0$ and in the presence 
of hyperons, we do not see antikaon condensation even up to a density 
of around 10 $\rho_0$.

Contrary to the case of neutron stars, the temperature in proto-neutron 
stars cannot be considered to be zero. In this case, we include the heat
bath of hadronic and quark quasiparticles within the grand canonical 
potential of the system. Also, instead of having a constant temperature 
throughout the star, we fix the entropy per baryon to $S/B=2$ allowing 
the temperature to increase from zero in the edge to $50$ MeV in the center
of the star following Ref. \cite{dexschr8}. This assumption is more realistic
and in agreement with dynamical simulations of star evolution and cooling 
\cite{Stein:2005nt,Pons:1998mm,Pons:2000xf,Pons:2001ar}. In Fig. 3 we plot
the kaon and antikaon energies at zero momentum for the nucleonic and 
hyperonic matter for the case of finite entropy per baryon. In this case, 
it is seen that for charge neutral nucleonic matter the onset of $K^-$ 
condensation is at a density of around 6 $\rho_0$. We might note that
in an earlier calculation \cite{banik08}, a temperature of around 
30 MeV is enough to melt the anti-kaon condensation in nucleonic matter.
With the inclusion of hyperons, our calculations show that
there is $K^-$ condensation at a density of around 7.4 $\rho_0$
as can be seen from Fig. 3.

Putting together these two features we can simulate the environment 
existent in proto-neutron stars. In Fig. 4 we plot $K$ and $\bar K$ 
energies at zero momentum in neutral matter with entropy per baryon 
$S/B=2$ and finite lepton number $Y_{L_e}=0.4$. For nucleonic matter, 
our model predicts $K^-$ condensation at a density of about 10 $\rho_0$, 
whereas the inclusion of hyperons does not give a possibility of $K^-$ 
condensation even up to a density of 10 $\rho_0$. These results show that 
there is no possibility for antikaon condensation in the earlier moments 
of the neutron star evolution independent of the composition considered 
to exist inside the star.

In the present model, the energy of $\bar {K^0}$ at zero momentum is never
zero and hence the condition for $\bar {K^0}$ condensation, i.e., 
$\mu_{\bar {K^0}}$=0 \cite{deba} is never met. In other words, 
the condensation of $\bar {K^0}$ does not take place at any moment
of the neutron star evolution in the our calculations using the SU(3) 
chiral model.  

\section{Summary and Conclusions}

To summarize, within a chiral SU(3) model we have investigated the density 
dependence of $K$, $\bar K$-meson energies in (proto-) neutron star bulk matter.
The model parameters are fitted to reproduce hadron masses in vacuum, nuclear
matter saturation properties, hyperon optical potentials and low energy KN 
scattering data. The model takes into account all terms up to the next to 
leading order arising in the chiral perturbative expansion for the
interactions of $K (\bar K)$-mesons with baryons. The medium modification 
of antikaons due to isospin asymmetry in dense matter can lead to the onset 
of antikaon condensation in the bulk charge neutral matter in 
(proto-) neutron stars.

We concluded that the drop in the antikaon mass in the medium leads 
to a possibility of antikaon condensation in cold neutron star matter. 
Our calculations show that in this case the onset of $K^-$ condensation 
takes place at a density of about 5.3 $\rho_0$ when the charge neutral 
matter comprises of only neutrons, protons, electrons and muons. This 
threshold density is seen to shift to higher densities of around 6.1 
$\rho_0$ when hyperons are included. The effect of neutrino trapping 
as well as finite temperature/entropy is a shift of the threshold density 
for the antikaon condensation in proto-neutron star matter. In this sense 
we conclude that there is no possibility for anti-kaon condensation 
in proto-neutron stars. Only after about a minute from the supernova 
explosion, when the star cools down via mainly neutrino emission the 
condensation can take place. Still, even in this case the anti-kaon 
condensation happens only in the very center of the star and does not 
cause considerable changes in macroscopic properties of the star, 
such as mass and radius.

\begin{acknowledgements}

One of the authors (AM) is grateful to the Institut
f\"ur Theoretische Physik Frankfurt for the
warm hospitality where the present work was initiated.
AM acknowledges financial support from Alexander von Humboldt 
Stiftung and from Department of Science and Technology, Government 
of India (project no. SR/S2/HEP-21/2006).
The use of the resources of the
Frankfurt Center for Scientific Computing (CSC) is additionally
gratefully acknowledged.
\end{acknowledgements}

\end{document}